\documentclass[twocolumn,twoside,slac_two]{revtex4}
\usepackage{graphicx}
\usepackage{fancyhdr}
\begin{document}

\title{Superconducting RF cavity R\&D for future accelerators}

%

\author{C.~M. Ginsburg}
\affiliation{Fermilab, Batavia, IL 60510 USA}

\begin{abstract}
High-beta superconducting radiofrequency (SRF) elliptical cavities are being
developed for several accelerator projects including Project X,
the European XFEL, and the International Linear Collider (ILC).  Fermilab
has recently established an extensive infrastructure for SRF cavity
R\&D for future accelerators, including cavity surface processing
and testing and cavity assembly into cryomodules.  Some highlights of
the global effort in SRF R\&D toward improving cavity performance,
and Fermilab SRF cavity R\&D in the context of global projects are reviewed.
\end{abstract}

\maketitle

\thispagestyle{fancy}


\section{Introduction}

Substantial effort is being expended in the quest for high gradients in
superconducting radiofrequency (SRF) cavities for ongoing and proposed projects: 
(1) the test/user facilities STF (KEK), NML (Fermilab), and FLASH (DESY), 
(2) the European XFEL currently under construction, (3) Project X at Fermilab, 
and (4) the International Linear Collider (ILC).  The common requirements or choices 
for the high-gradient cavities in these projects are gradients of at least 23 MV/m, 
$\beta$=1, elliptical shape and an accelerating mode frequency of 1.3 GHz.  Recent 
SRF R\&D highlights in the pursuit of high cavity gradients and the status
of Fermilab SRF infrastructure development are reviewed.

\section{Achieving High Gradient}

Achieving high gradient in SRF cavities for current projects requires pure niobium, 
an excellent surface quality, and a good geometry.

For the ILC~\cite{ilc_rdr}, the niobium sheets from which cavities are formed must have a 
residual resistivity ratio (RRR) of at least 300, consistent with very high purity niobium, 
to ensure good superconducting properties.

\begin{figure}[h]
\centering
\includegraphics[width=80mm,angle=0]{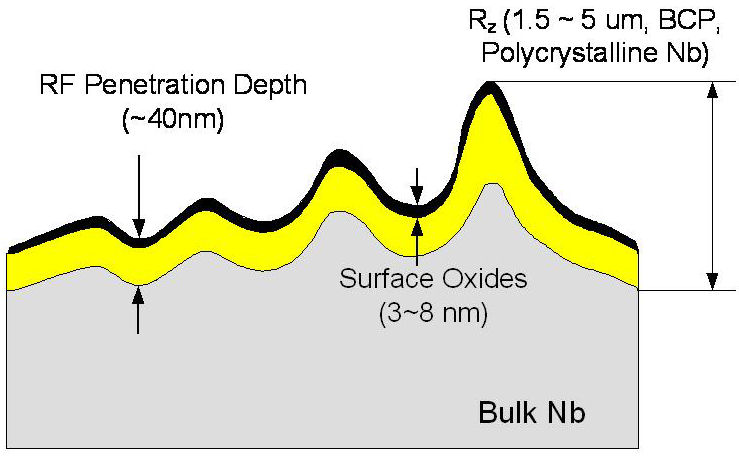}
\caption{Simplified drawing of an SRF cavity surface. [Courtesy of Hui Tian.]} \label{fig_surface}
\end{figure}

Because RF fields occupy the first $\sim$40 nm of the inner cavity surface for these cavities, 
as shown in Fig.~\ref{fig_surface}, the quality of the innermost surface is critical 
and must be carefully controlled both during cavity fabrication and test preparation.
To avoid the inclusion of contaminants or defects apparent after the sheet manufacturing,
eddy current scanning (ECS) of all sheets is performed before cavity 
fabrication.  ECS has proven to be a very useful feedback mechanism for 
material vendors, constributing substantially to an overall material quality improvement 
in recent years.  
After fabrication, the cavity inner surface must be very smooth, with no inclusion of 
foreign particles, or topological defects such as bumps or pits or sharp niobium grain boundaries,
which could increase surface resistance locally.
In addition to the strict quality assurance for the material, the introduction of dust or other 
microscopic contaminants must be avoided after the final surface preparation, to prevent
field emission which can become dark current during machine operation and cause undesirable
heating in superconducting elements and damage to accelerator components.

In selecting a cavity geometry for optimum performance, cavity cell shapes are typically
optimized for low peak surface magnetic field (H$_{\rm peak}$) and 
low peak surface electric field (E$_{\rm peak}$) relative to the gradient 
(E$_{\rm acc}$), as well as ease of surface processing and fabrication.

\begin{figure}[h]
\centering
\includegraphics[width=80mm,angle=0]{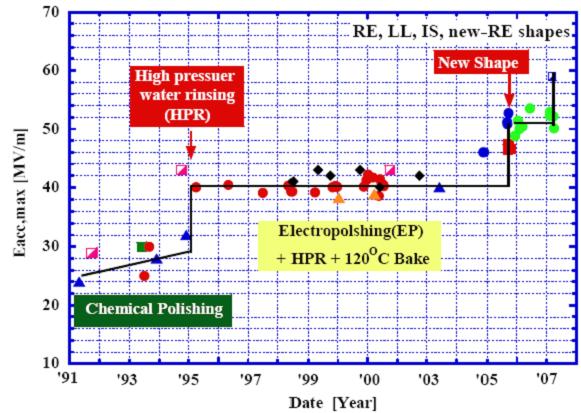}
\includegraphics[width=80mm,angle=0]{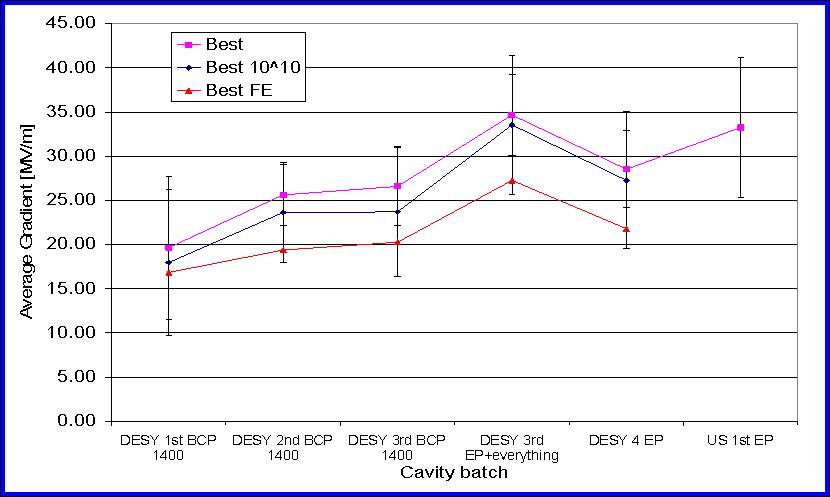}
\caption{Maximum cavity gradient as a function of time for 1-cell
cavities of non-standard shapes (top, courtesy of Kenji Saito) and of 9-cell cavities of
Tesla shape (bottom, courtesy of Lutz Lilje). All tests were performed at 2K. 
(Data are from KEK, Cornell, JLab, and DESY.)} \label{fig_SRF2007_Saito}
\end{figure}

Many cavities have reached 35 MV/m or more in the last decade~\cite{SRF2007_Saito,MAC2007_Lilje}, 
particularly single-cell cavities of varying elliptical shapes
and 9-cell Tesla-shape cavities, as shown in Fig.~\ref{fig_SRF2007_Saito}.

\section{Surface Processing}
The surface treatment intended to maximize cavity performance, developed for the 
ILC~\cite{TTC_surfaceprep}, includes initial preparation steps to remove $\sim$$150 \ \mu$m from 
the inner surface using electropolishing (EP).  
This initial removal step may be performed with centrifugal barrel 
polishing (CBP) or buffered chemical polishing (BCP) at some labs; however, the maximum 
gradients reached with BCP as the initial preparation step are lower than those achieved 
using the other methods.  Cavities then undergo an 800°C annealing step, to drive hydrogen 
from the surface.  The final preparation steps include degreasing with detergent, another 
light electropolishing ($\sim$$20 \ \mu$m), a high pressure rinse (HPR) with ultrapure water, drying 
in a class-10 cleanroom, and then evacuation and low-temperature baking (120°C) for about 
48 hours after the final assembly with couplers.  Additional surface treatments to address
field emission, which occur after the final EP, will be described later.
The primary methods for material removal during surface preparation are CBP, EP and BCP.  

\begin{figure}[h]
\centering
\includegraphics[width=80mm,angle=0]{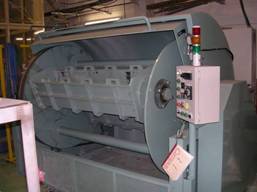}
\includegraphics[width=80mm,angle=0]{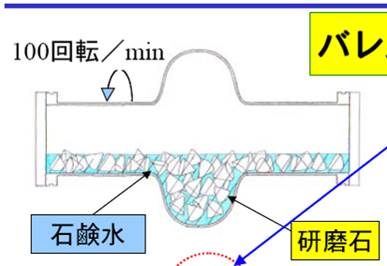}
\caption{The KEK CBP machine (top) and a diagram showing the
CBP process (bottom).  [Courtesy of Kenji Saito.]} 
\label{fig_CBP}
\end{figure}

CBP is a standard technique developed for cavities at KEK in which abrasive small stones 
are placed into a cavity with water to form a slurry and the cavity is rotated.  The KEK
CBP machine and a diagram showing the CBP process are shown in Fig.~\ref{fig_CBP}.  
As a centrifugal process, material is preferentially removed from the equator region.  
Since standard cavities have an equator weld, CBP is very effective in smoothing the weld.  

\begin{figure}[h]
\centering
\includegraphics[width=80mm,angle=0]{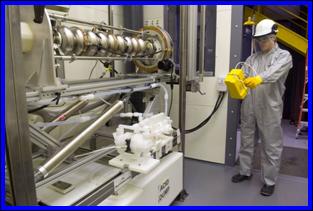}
\includegraphics[width=80mm,angle=0]{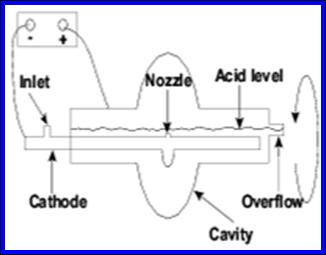}
\caption{The ANL EP machine (top, courtesy of Michael Kelly),
and a diagram showing the EP process (bottom, courtesy of Lutz Lilje).} 
\label{fig_EP}
\end{figure}

EP is an electrolytic current-supported material removal, and has been developed for use 
on cavities by KEK in collaboration with industrial partners, and adopted at most labs.  
The EP process is complementary to CBP because the material removal is preferentially on the 
iris.  
The ANL EP machine and a diagram showing the EP process are shown in Fig.~\ref{fig_EP}.
In this case, the niobium cavity functions as an anode, and an aluminum cathode is inserted 
on the cavity axis.  
The electrolyte is HF(40\%):H$_2$SO$_4$ in a ratio of 1:9 by volume.  Some sulfur 
remains on the surface after EP and will cause field emission unless it is removed.  

BCP involves filling a cavity with a combination of 
HF(40\%):HNO$_3$(65\%):H$_3$PO$_4$(85\%) in a 1:1:2 ratio by volume.  BCP has been shown to cause hydrogen 
contamination at the surface; this problem can be mitigated by using the appropriate proportion 
of acid and buffer and by keeping the temperature below about 15°C.  BCP is rather less 
expensive than EP and is often sufficient to produce cavities attaining gradients as high 
as about 25 MV/m.  However, it tends to enhance grain boundaries which may degrade the 
performance of standard fine-grain cavities.  

\begin{figure}[h]
\centering
\includegraphics[width=80mm,angle=0]{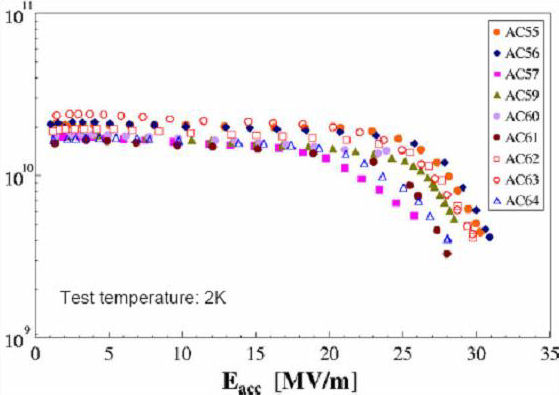}
\includegraphics[width=80mm,angle=0]{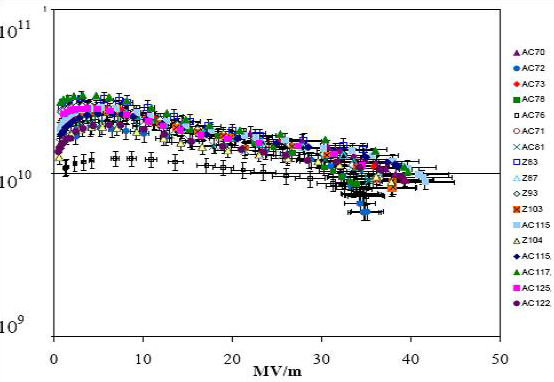}
\caption{Comparison of cavities processed with BCP (top) and
EP (bottom).  The EP'd cavities reach noticeably higher gradients.
[Courtesy of DESY.]} 
\label{fig_DESY_Prod3}
\end{figure}

\begin{figure}[h]
\centering
\includegraphics[width=80mm,angle=0]{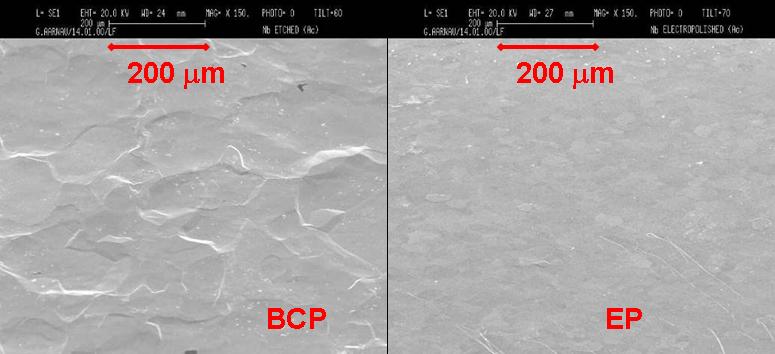}
\caption{Comparison of a sample surface after BCP (left) or EP (right).
The EP'd surface is noticeably smoother than the BCP'd surface.
[Courtesy of DESY.]} 
\label{fig_BCP_vs._EP}
\end{figure}

An improvement to the average maximum gradient was seen after the previous BCP standard was 
replaced with EP, as shown in Fig.~\ref{fig_SRF2007_Saito}.
The comparison of best test results of DESY cavities treated with BCP and EP is shown in 
Fig.~\ref{fig_DESY_Prod3}, revealing that the maximum gradient achieved with BCP is
$\sim$25-30 MV/m, whereas the maximum gradient achieved with EP is rather higher at $\sim$35-40 MV/m.
A comparison of the resulting surface smoothness is shown in Fig.~\ref{fig_BCP_vs._EP}.
Although it is possible now to achieve gradients in 9-cell cavities higher than 35 MV/m, 
there are several factors which limit cavity performance.

\section{Reducing Field Emission}

\begin{figure}[h]
\centering
\includegraphics[width=40mm,angle=0]{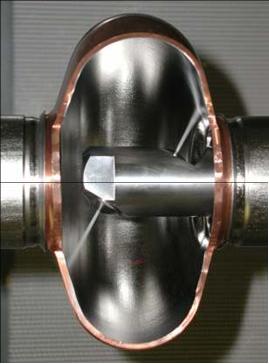}
\caption{The DESY dry ice nozzle. [Courtesy of DESY.]} 
\label{fig_dryice_nozzle}
\end{figure}

One of the factors which limits cavity performance is the presence of field emission.
Four recent field emission studies have been performed: flash EP (also known as fresh EP), 
dry ice cleaning, degreasing, and ethanol rinsing. 
In tests at KEK using six single-cell Ichiro-shape cavities, a 3 $\mu$m EP using fresh acid 
after the final EP was studied~\cite{flashEP}.  A gradient improvement to both average and rms was 
observed.  Furthermore, the treatment was found to increase the gradient at which field 
emission turns on.
Dry ice cleaning is a less developed but promising alternative being developed at DESY~\cite{dry_ice}.  
It reduces or eliminates contaminating particles in several ways: they become brittle upon 
rapid cooling, they encounter pressure and shearing forces as CO$_2$ crystals hit the surface, 
and they are rinsed due to the 500 times increased volume after sublimation.  Also, LCO$_2$ 
is a good solvent and detergent for hydrocarbons and silicones, etc.  Dry ice cleaning is 
a dry process which leaves no residues, because the loosened contaminants are blown out 
the ends of the cavities by the positive pressure, and can be performed in a horizontal 
orientation. The DESY dry ice nozzle is shown in Fig.~\ref{fig_dryice_nozzle}.
Dry ice cavity cleaning might be possible after coupler installation, 
a procedure which risks introducing field emission.  Improved field emission characteristics 
have been seen in single-cell cavity tests, and an extension of the system to 9-cell 
cavities is planned.

One of several degreasing R\&D studies is that of a KEK 9-cell Ichiro-shape cavity which 
was processed and tested at KEK and JLab~\cite{geng_detergent}.  In an initial test at JLab, the cavity 
produced substantial field emission which was observed at gradients above 15 MV/m.  
After the cavity was ultrasonically cleaned with a 2\% Micro-90 solution and standard HPR, 
the field emission was substantially reduced or even eliminated.

An extensive study of ethanol rinsing after the final EP was carried out at DESY~\cite{DESY_ethanol}.  
Out of 33 cavities used in the study, 20 were treated using the standard procedure and 
13 with an additional ethanol rinse after the final EP.  The number of tests in which 
field emission was observed was substantially reduced.  In addition, the maximum gradient 
was somewhat improved, from $27 \pm 4$ MV/m without ethanol to $31 \pm 5$ MV/m with ethanol rinse.
Since field emission was found to be reduced so substantially, 
the ethanol rinse is now part of the standard DESY cavity treatment.

\section{Manufacturing and Shape Studies}

The vast majority of cavities which have been processed and tested in the last decade 
are Tesla-shape fine-grain cavities~\cite{TESLA_cavities}.  Some fundamental changes to the standard 
Tesla-shape fine-grain cavities which have been under investigation include changes 
to the fabrication techniques, material composition and cavity shape.

Because electron-beam welding is a substantial portion of the cavity fabrication cost, and may 
be a source of surface defects which cause premature quenches (described later), reducing the 
number of equator welds required in cavity fabrication would be advantageous.
Recently, a 9-cell cavity was fabricated at DESY using a hydroforming technique~\cite{Singer_hydroforming}.  
The cavity was 
built from three 3-cell hydroformed units, so it had only two iris welds and two beampipe welds.  
The surface was treated with the standard procedure including ethanol rinse.  The cavity 
performance was comparable to cavities built using standard techniques.

Material is wasted, and a lot of time is required, to roll sheets and stamp out the 
disks from which fine-grain cavities are made.  In principle, it should be possible to 
save manufacturing costs by slicing an ingot directly into large-grain sheets. It may also 
be possible to achieve high gradients using BCP only, since the performance of large-grain 
cavities should be less sensitive to the grain boundary enhancement seen in fine-grain 
cavities, as long as the grain boundaries are strategically located in low surface-field 
regions.  
Recent experience at DESY with large-grain cavities shows their performance is comparable to 
fine-grain cavities.  It is still unclear whether BCP will be sufficient or whether EP will 
be necessary.  Many 1-cell tests have shown high gradients, in a range comparable to 
fine-grain cavities.  Recent tests on three 9-cell large-grain cavities~\cite{Singer_LG} showed gradients 
comparable to fine-grain cavities as well.  Effective large-grain ingot cutting methods are being
pursued in industry.
Single-crystal niobium cavities have been difficult to produce, because it is difficult to 
produce large diameter single-grain ingots.  Six single-cell single-grain cavities of varying 
shape and fabrication technique were fabricated, processed and tested recently by a JLab/DESY 
collaboration~\cite{Kneisel_singlecrystal}, with performance found to be comparable to that 
of fine-grain cavities.  
Unless substantial performance improvement is seen, the difficulty of producing single-grain cavities 
may not justify the effort required.
In both the large-grain and single-grain cases, further study of crystal orientation effects 
is needed.  For a summary of recent large-grain and single-grain cavity work, see Ref.~\cite{Kneisel_LG}. 

It may be possible to increase the RF breakdown magnetic field of superconducting cavities by
creating a multilayer coating of alternating insulating layers and thin superconducting 
layers~\cite{Gurevich}.  By using multilayers thinner than the RF penetration depth, the critical magnetic 
field can be increased from niobium H$_{\rm c1}$ to one similar to high-T$_{\rm c}$ superconductors, thereby 
significantly improving the maximum gradient achievable.
Recently, a cavity was prepared with such a composite surface and tested~\cite{Proslier}.  
A 10 nm layer of Al$_2$O$_3$ was chemically bonded to the niobium surface of an existing 
single-cell cavity using atomic layer deposition.  
The surface was then covered by a  3 nm layer of Nb$_2$O$_5$.  The cavity showed
promising early results, and further study is underway.

\begin{figure}[h]
\centering
\includegraphics[width=50mm,angle=0]{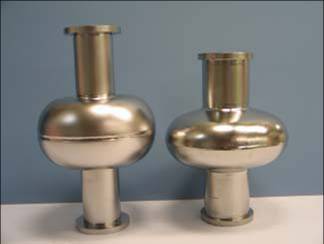}
\includegraphics[width=30mm,angle=0]{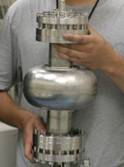}
\caption{Single-cell 1.3 GHz elliptical cavities: Cornell re-entrant (left),
Tesla (center) and Ichiro (right).} 
\label{fig_shapes}
\end{figure}

\begin{figure}[h]
\centering
\includegraphics[width=80mm,angle=0]{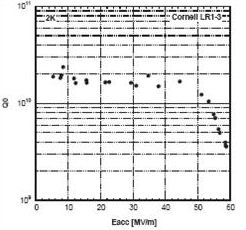}
\caption{The record SRF 1-cell cavity gradient. [Courtesy of Cornell University.]} 
\label{fig_Cornell_record}
\end{figure}

Cell accelerating length and equator diameter are fixed by $\beta$ and frequency respectively.  
However, the details of the shape may be optimized for low field emission 
(low E$_{\rm peak}$/E$_{\rm acc}$) 
and reduced sensitivity to the fundamental maximum surface magnetic field 
(low H$_{\rm peak}$/E$_{\rm acc}$); 
see, e.g., Ref.~\cite{Jacek_shapes}.
Excellent results on single-cell elliptical cavities have recently been obtained. 
A comparison of three shapes of elliptical single-cell cavities is shown in Fig.~\ref{fig_shapes}.
A re-entrant shape cavity built and tested by Cornell University recently reached 
59 MV/m~\cite{Cornell_record}, as shown in Fig.~\ref{fig_Cornell_record},
setting a world record for the type of cavities described in this paper.  Excellent results 
have also been achieved with an Ichiro shape cavity at KEK, with a record of 
53.5 MV/m~\cite{SRF2007_Saito}.  
Furthermore, $46.7 \pm 1.9$ MV/m was reached on six single-cell cavities with optimized surface 
treatment parameters~\cite{flashEP}.  
Another low-loss shape single-cell cavity which was processed 
and tested by a DESY/KEK collaboration reached 47.3 MV/m~\cite{Furuta_EPAC06}.
It is rather more difficult to manufacture an excellent multi-cell cavity than an excellent 
single-cell cavity, and the very high gradients seen in single-cell alternative-shape cavities
have not been achieved in 9-cell versions yet.  One 9-cell Ichiro-shape cavity, without 
endgroups, which was processed and tested by a KEK/JLab collaboration reached 
up to 36 MV/m~\cite{Geng_LINAC08}.  

\section{Understanding Cavity Behavior}

Quenches and field emission appear as hot spots on the outer cavity surface.  Temperature 
mapping (T-map) systems have been used at nearly all of the labs active in this field for many 
years to study these phenomena~\cite{Padamsee}. The existing T-map systems primarily use Allen-Bradley 
carbon resistors as thermal sensors, as developed at Cornell~\cite{Cornell_thermometry}.  
T-map systems commonly in use for many years at DESY include a fixed type for single-cell cavities with 
768 sensors~\cite{DESY_fixed_thermometry}, and a rotating system for 9-cell cavities with 
128 sensors~\cite{DESY_rotating_thermometry}.  
A fixed thermometry system for 9-cell cavities with four sensors around each equator 
and a few sensors on the endgroups has been developed at KEK and was used for tests of 
STF Tesla-like cavities~\cite{KEK_diagnostics}.  Second sound detection has been used at ANL for quench 
location on split-ring resonators~\cite{Shepard_secondsound}.  
A selection of new hot spot detection systems 
are described; these systems vary in coverage, flexibility, and in the number of cavity 
tests required to extract useful T-map information.

\begin{figure}[h]
\centering
\includegraphics[width=80mm,angle=0]{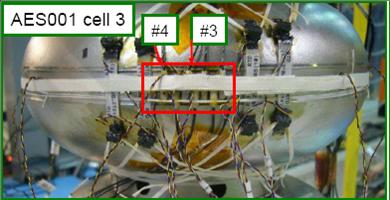}
\caption{Fermilab Cernox thermometry system installed on a 9-cell cavity
in the region of a pre-determined hot spot.} 
\label{fig_fast_thermometry}
\end{figure}

A system of Cernox temperature sensors has been developed at Fermilab~\cite{fast_thermometry} 
and used for quench location.  
Up to 32 sensors may be attached as needed to suspect locations; therefore 
the system is very flexible but also time consuming to install, typically requiring several 
cavity test cycles to conclusively locate quenches.  It is highly portable and suitable for
any cavity shape.  An example installation of this system on a 9-cell cavity is shown
in Fig.~\ref{fig_fast_thermometry}, in which a hard quench at 16 MV/m was observed, 
accompanied by a temperature rise of about 100 mK, above the 2K operation, over about 
2 sec in sensors \#3
and \#4 before the quench.  The quench was seen on all sensors.

New multi-cell T-map systems have also been developed.  A new 2-cell system at JLab was 
recently commissioned~\cite{JLab_2-cell_thermometry}, with 160 Allen-Bradley sensors 
installed around each of the two equators.  
Understanding cavity behavior requires two cooldowns: one to measure the 
TM010 passband modes and determine within two the limiting cells, then a second after 
the T-map installation to measure the temperature distribution of the limiting cell.  
At LANL, a new 9-cell T-map system has been developed~\cite{LANL_9-cell_thermometry} 
which employs 4608 Allen-Bradley 
sensors and a multiplexing scheme to map an entire 9-cell cavity in a single test yet 
limit the number of cables leaving the cryostat.  The preliminary results are very promising.  
A 9-cell T-map system at Fermilab using 8640 diodes as a multiplexed thermometer array 
is under development~\cite{diode_thermometry}.  
All of these multi-cell T-maps are specifically designed for the Tesla cavity shape. 

\begin{figure}[h]
\centering
\includegraphics[width=80mm,angle=0]{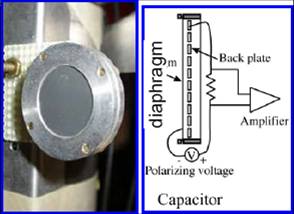}
\caption{The Cornell second sound sensor (left, courtesy of Cornell University) 
and a diagram showing the function of the sensor (right, courtesy of Genfa Wu).} 
\label{fig_secondsound_detector}
\end{figure}

\begin{figure}[h]
\centering
\includegraphics[width=80mm,angle=0]{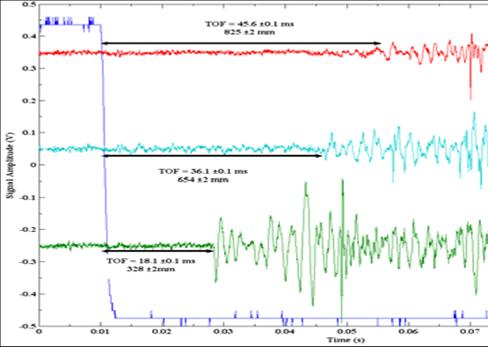}
\caption{Cornell data showing the arrival of second sound waves at three
different sensors. [Courtesy of Zachary Conway.]} 
\label{fig_secondsound_data}
\end{figure}

Cornell has developed a new quench location system using second sound sensors~\cite{Cornell_secondsound}.  
Second sound is a thermal wave which can propagate only in superfluid helium.  It is generated when 
a heat pulse is transmitted from a heat source, such as a quench, through superfluid helium.  
In this system, eight sensors detect the arrival of the temperature oscillation, and the 
location is determined from the relative timing of the arrival of the oscillation from 
different sensors.  This simple system is suitable for any cavity shape or number of cells 
and can locate quenches in a single cavity test.  The detector is shown in 
Fig.~\ref{fig_secondsound_detector}, and an example of the data used to triangulate the
location of the second sound source is shown in Fig.~\ref{fig_secondsound_data}.

\begin{figure}[h]
\centering
\includegraphics[width=80mm,angle=0]{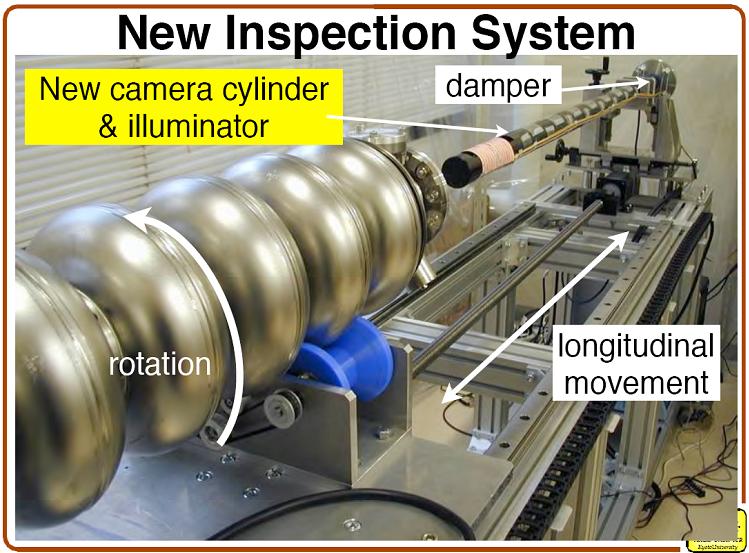}
\caption{The KEK/Kyoto University cavity inspection camera system.
[Courtesy of Yoshihisa Iwashita.]} 
\label{fig_kyoto_camera}
\end{figure}

A resurgence of interest in optical inspection occurred recently when several 
cavities with hard quench limitation in the 15-20 MV/m range, were observed to have surface 
defects correlated with hot spots, using a new optical inspection system developed by 
Kyoto University/KEK with a clever lighting technique and excellent resolution.
This optical system~\cite{Kyoto_camera} consists of an integrated camera, mirror, and lighting system on 
a fixed rod; the cavity is moved longitudinally and axially with respect to the camera system.  
The lighting is provided by a series of electroluminescent strips, which provide lighting with 
low glare on the highly polished surface.  These strips can be turned off and on, and the 
resulting shadows studied for an effective 3D image of the depth or height of the defect.  
The camera resolution is about 7 $\mu$m.  A photograph of the Kyoto/KEK camera system
is shown in Fig.~\ref{fig_kyoto_camera}.
An optical inspection system employed regularly at JLab~\cite{Geng_LINAC08} includes 
a Questar long-distance microscope and mirror to inspect any cavity inner surface.  
This system uses electroluminescent lighting.
Additional optical inspection systems are in use or under development at other labs.

Because many of the cavity-limiting surface defects found by thermometry and optical 
inspection are located in the heat-affected zone of the electron-beam welds, weld properties 
are under intense scrutiny.  Understanding which weld parameters might contribute to creating 
such defects, and ultimately eliminating the defect occurrence, is critical to improving 
the yield of high-gradient cavities.  Note that cavities with such defects are largely,
but not exclusively, from new cavity manufacturers.
One group has attempted to reproduce these features on samples~\cite{Cooley},  
to permit their systematic study. In time, study of cavity surface quality may improve cavity
quality in a manner similar to the improvement seen in niobium sheet quality resulting from 
regular eddy current scanning.

\section{Fermilab Infrastructure for SRF Development}

At Fermilab, the SRF infrastructure has been substantially built up
in recent years.  The key program goals are the achievement of
high gradients with high yields for future accelerators, the
testing of cavities, and the ramp up of the production rates needed 
to construct cryomodules for Project X and ILC R\&D.
The infrastructure which has been built to achieve these
goals includes facilities for single-cell cavity R\&D, an ANL/FNAL
joint cavity processing facility, a vertical cavity test facility,
a horizontal test stand, a cryomodule assembly facility, and 
a cryomodule test facility.

\begin{figure}[h]
\centering
\includegraphics[width=40mm,angle=0]{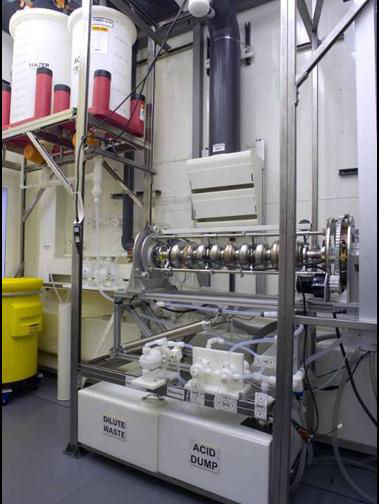}
\includegraphics[width=40mm,angle=0]{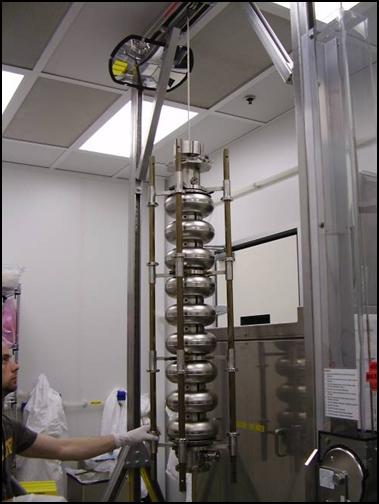}
\includegraphics[width=40mm,angle=0]{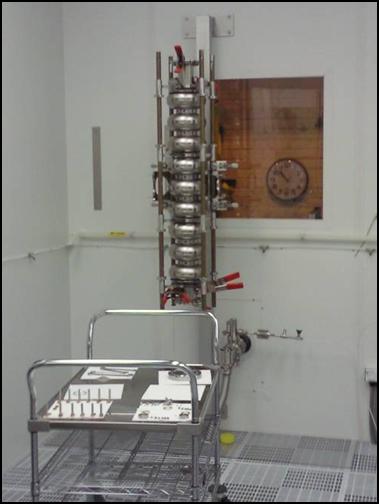}
\includegraphics[width=40mm,angle=0]{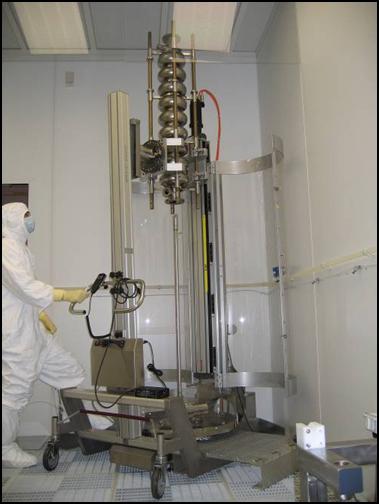}
\caption{ANL/FNAL join facility infrastructure: (clockwise from top left)
EP, ultrasonic degreasing, HPR and assembly stands. [Courtesy of Michael Kelly and Damon Bice.]} 
\label{fig_ANL-CPF}
\end{figure}

The ANL/FNAL cavity processing facility provides surface processing
and assembly of cavities for vertical test, for  
both 1-cell and 9-cell elliptical cavities.
Recent photographs of the primary facility components: EP, ultrasonic degreasing, 
high-pressure ultrapure water rinsing, and assembly support and vacuum 
leak testing, are shown in Fig.~\ref{fig_ANL-CPF}.  All of these 
components are now in routine operation.

\begin{figure}[h]
\centering
\includegraphics[width=80mm,angle=0]{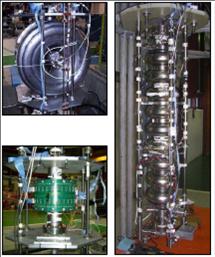}
\caption{Some of the cavities recently tested at the Fermilab VCTF:
(clockwise from top left) SSR1, 9-cell Tesla-shape, and 1-cell Tesla-shape 
(with diode thermometry) cavities. [Courtesy of Bill Mumper.]} 
\label{fig_VCTF}
\end{figure}

\begin{figure}[h]
\centering
\includegraphics[width=80mm,angle=0]{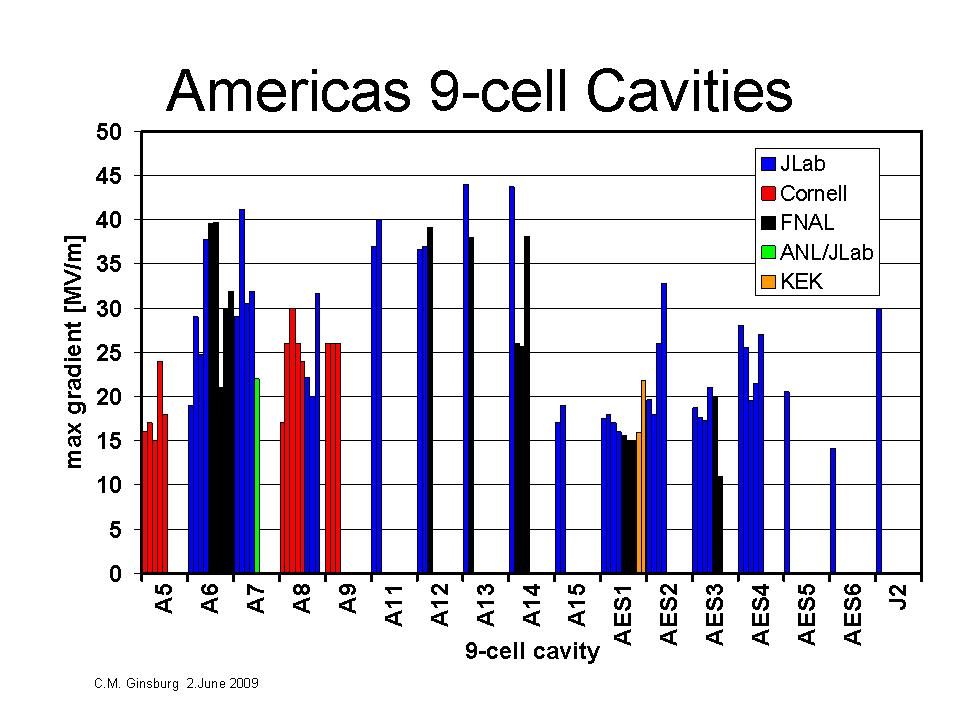}
\caption{A summary of maximum gradients from 9-cell vertical tests in the ILC Americas
region.  The cavities tested at the VCTF are shown by the black histogram.} 
\label{fig_AmericasCavitySummary}
\end{figure}

At the FNAL vertical cavity test facility (VCTF), more than 40 cavity
tests have been performed in FY08/FY09, where a test is defined as a cryogenic cycle.  The
test cavities have been primarily 9-cell and single-cell 1.3 GHz elliptical
cavities, with two 325 MHz single-spoke resonators included.
The tests have primarily been for instrumentation development, e.g.,
variable input coupler, thermometry, cavity vacuum pumping system, 
and for cavity diagnostic tests for vendor development.  Upgrades are planned
to increase the cavity test throughput to more than 200 cavity tests
per year by October 2011, to support Project X and ILC R\&D.  Some of
the cavities recently tested at the VCTF are shown in Fig.~\ref{fig_VCTF}.
A summary of the vertical test results of 9-cell cavities is shown in 
Fig.~\ref{fig_AmericasCavitySummary}.

\begin{figure}[h]
\centering
\includegraphics[width=80mm,angle=0]{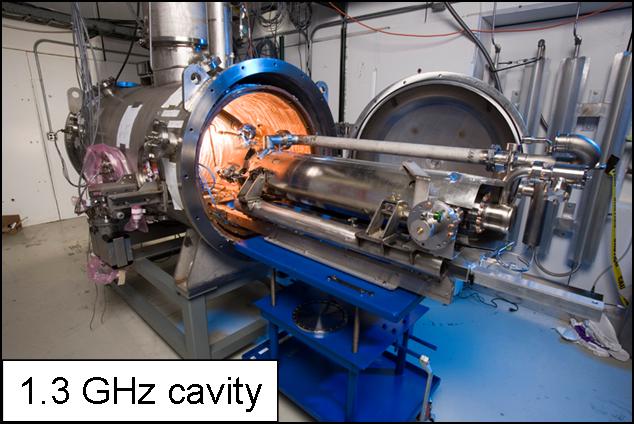}
\caption{A photograph of the HTS, showing a 9-cell 1.3 GHz cavity
being installed. [Courtesy of Fermilab Visual Media Services.]} 
\label{fig_HTS}
\end{figure}

The horizontal test stand (HTS), which is used to test 9-cell cavities which
have been welded into a helium tank, i.e., dressed, using pulsed high power, is now operational.
It has been commissioned for both 1.3 and 3.9 GHz cavities.  Four 3.9 GHz cavities were
tested in 2008; they have been installed in a cryomodule which is currently at DESY.
Upgrades are planned to expand the capacity with the addition of a second
cryostat by 2012.  A photograph of the HTS is shown in Fig.~\ref{fig_HTS}.

\begin{figure}[h]
\centering
\includegraphics[width=40mm,angle=0]{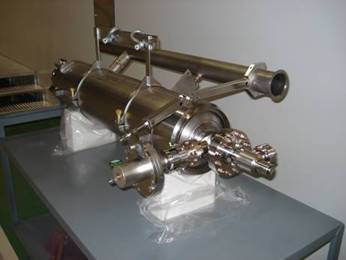}
\includegraphics[width=40mm,angle=0]{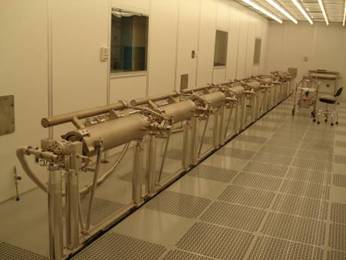}
\includegraphics[width=40mm,angle=0]{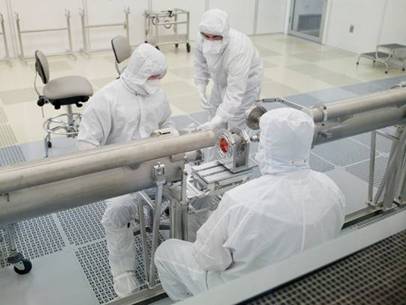}
\includegraphics[width=40mm,angle=0]{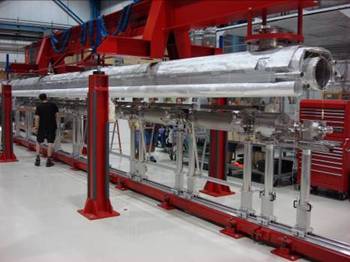}
\includegraphics[width=40mm,angle=0]{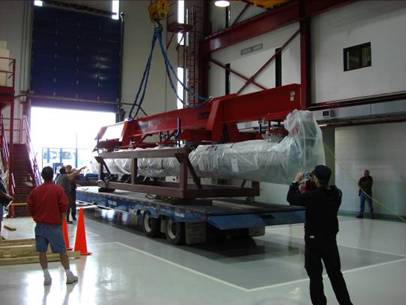}
\caption{Recent photographs of the CAF-MP9 facility, showing stages in the
assembly of the first cryomodule, CM1. [Courtesy of Tug Arkan.]} 
\label{fig_CAF-MP9}
\end{figure}

\begin{figure}[h]
\centering
\includegraphics[width=40mm,angle=0]{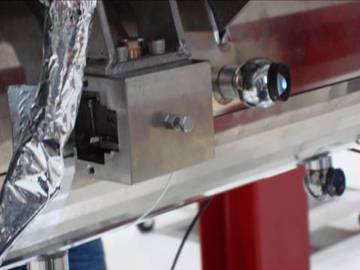}
\includegraphics[width=40mm,angle=0]{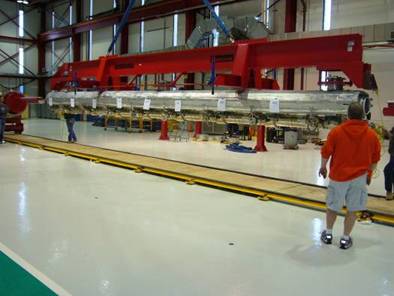}
\includegraphics[width=40mm,angle=0]{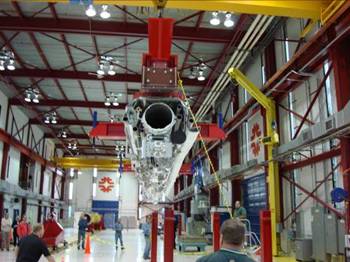}
\includegraphics[width=40mm,angle=0]{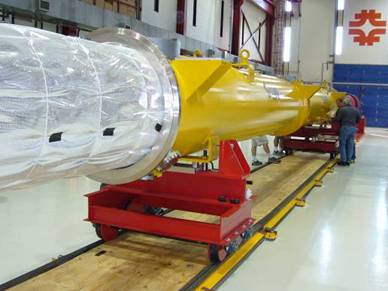}
\includegraphics[width=40mm,angle=0]{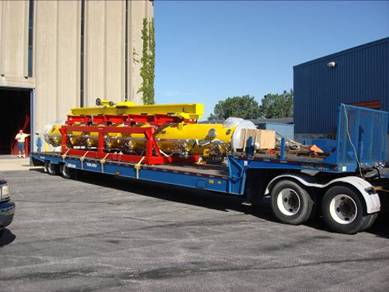}
\caption{Recent photographs of the CAF-ICB facility, showing stages in the
assembly of the first cryomodule, CM1.  [Courtesy of Tug Arkan.]} 
\label{fig_CAF-ICB}
\end{figure}

Two cryomodule assembly facilities (CAF's) have been built at Fermilab, CAF-MP9
and CAF-ICB.  The CAF-MP9 facility receives dressed cavities and their
peripheral parts, and assembles these into a cavity string in a class-10
cleanroom.  Then, the string assembly is installed onto the cold mass and
transported to CAF-ICB.  At CAF-ICB, the cavity string is aligned and assembled
into the vacuum vessel.  After the cryomodule is complete, it is shipped to
the cryomodule test facility (NML) for testing.  One cryomodule, CM1, built 
with cavities from DESY, has been assembled in the CAF facilities.  Recent photographs
of the CAF-MP9 and CAF-ICB, showing steps in the assembly of CM1,
are shown in Figs.~\ref{fig_CAF-MP9} and \ref{fig_CAF-ICB}, respectively.

\begin{figure}[h]
\centering
\includegraphics[width=80mm,angle=0]{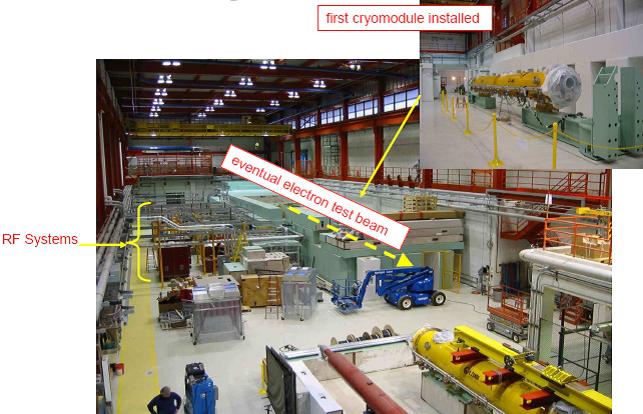}
\caption{A recent photograph of the NML facility, indicating where
the first cryomodule, CM1, has been installed. [Courtesy of Jerry Leibfritz.]} 
\label{fig_NML}
\end{figure}

The cryomodule test facility, NML, will test one RF
unit, i.e., 3 cryomodules, using a 10 MW RF system and an electron beam
with ILC parameters.  Various Project-X parameters will also be tested with beam
in this facility.  NML will provide the capability to conduct
advanced accelerator R\&D for future accelerator components.  Through the
end of FY09, the facility is being prepared to test the first Fermilab-assembled
cryomodule, CM1, without beam, requiring completion of the RF system and the
cryogenic system.  A recent photograph of the NML facility is shown in Fig.~\ref{fig_NML}.

\section{Conclusions and Outlook}

Highlights of the rich R\&D activity in the quest for high gradient and reduced cost have been 
described.  Very high gradients have been measured in niobium SRF cavities: more than 50 MV/m 
in single-cell cavities of various shapes, and more than 35 MV/m in several 9-cell Tesla-shape 
cavities.  Several promising new results from large-grain 9-cell cavities and hydroformed 
cavities address the cost and reliability issues of these cavities.  Several cavities which 
have been limited to 15-20 MV/m by hard quench have been studied using various techniques, 
and the outlook is good that useful information can be fed back to cavity manufacturers to 
improve the yield of high-gradient cavities over time.  Surface treatment is crucial for 
optimum performance, and several promising studies on final preparation methods have been 
found to sharply reduce field emission.  Fundamental SRF studies are also underway, many showing 
promising results.

An overview of Fermilab infrastructure in pursuit of SRF R\&D has been given.
Among the recent accomplishments, one 1.3 GHz cryomodule using DESY dressed cavities has been 
built and transported to the cryomodule test facility for testing.  Most
key infrastructure components are now in place, with final commissioning underway.

\begin{acknowledgments}
I am grateful to my colleagues for their contributions to the content 
for this talk, as referenced.  I especially thank those who graciously permitted me to 
show their unpublished data.  

This manuscript has been authored by Fermi Research Alliance, LLC under Contract No. 
DE-AC02-07CH11359 with the U.S. Department of Energy. The United States Government 
retains and the publisher, by accepting the article for publication, acknowledges 
that the United States Government retains a nonexclusive, paid-up, irrevocable, 
worldwide license to publish or reproduce the published form of this manuscript, 
or allow others to do so, for United States Government purposes.
\end{acknowledgments}

\bigskip 

\end{document}